\documentclass[conference]{IEEEtran}
\usepackage{fancyhdr}

\usepackage[acronyms,nonumberlist,nopostdot,nomain,nogroupskip]{glossaries}
\newacronym{6g}{6G}{sixth generation}
\newacronym{3gpp}{3GPP}{3rd Generation Partnership Project}
\newacronym{5g}{5G}{5th generation}
\newacronym{aimd}{AIMD}{Additive Increase Multiplicative Decrease}
\newacronym{am}{AM}{Acknowledged Mode}
\newacronym{gbps}{Gbps}{Gigabit-per-second}
\newacronym{amc}{AMC}{Adaptive Modulation and Coding}
\newacronym{aqm}{AQM}{Active Queue Management}
\newacronym{awgn}{AGWN}{Additive White Gaussian Noise}
\newacronym{balia}{BALIA}{Balanced Link Adaptation}
\newacronym{bdp}{BDP}{Bandwidth-Delay Product}
\newacronym[plural=\gls{cnn}s,firstplural=convolutional neural networks (CNNs)]{cnn}{CNN}{convolutional neural network}
\newacronym[plural=\gls{dnn}s,firstplural=deep neural networks (DNNs)]{dnn}{DNN}{deep neural network}
\newacronym{bf}{BF}{Beamforming}
\newacronym{b5g}{B5G}{\gls{5g} and beyond}
\newacronym{cc}{CC}{Congestion Control}
\newacronym{cdf}{CDF}{Cumulative Distribution Function}
\newacronym{cn}{CN}{Core Network}
\newacronym{cqi}{CQI}{Channel Quality Information}
\newacronym{cp}{CP}{Control Plane}
\newacronym{csirs}{CSI-RS}{Channel State Information - Reference Signal}
\newacronym{dc}{DC}{Dual Connectivity}
\newacronym{dce}{DCE}{Direct Code Execution}
\newacronym{dci}{DCI}{Downlink Control Information}
\newacronym{dl}{DL}{deep learning}
\newacronym{dmr}{DMR}{Deadline Miss Ratio}
\newacronym{dmrs}{DMRS}{DeModulation Reference Signal}
\newacronym{e2e}{E2E}{End-to-End}
\newacronym{ecn}{ECN}{Explicit Congestion Notification}
\newacronym{edf}{EDF}{Earliest Deadline First}
\newacronym{enb}{eNB}{evolved Node Base}
\newacronym{epc}{EPC}{Evolved Packet Core}
\newacronym{es}{ES}{Edge Server}
\newacronym{fdma}{FDMA}{Frequency Division Multiple Access}
\newacronym{fdd}{FDD}{Frequency Division Duplexing}
\newacronym[firstplural=Radio Access Technologies (RATs)]{rat}{RAT}{Radio Access Technology}
\newacronym{fs}{FS}{Fast Switching}
\newacronym{ftp}{FTP}{File Transfer Protocol}
\newacronym{gnb}{gNB}{Next Generation Node Base}
\newacronym{harq}{HARQ}{Hybrid Automatic Repeat reQuest}
\newacronym{hetnet}{HetNet}{Heterogeneous Network}
\newacronym{hh}{HH}{Hard Handover}
\newacronym{hol}{HOL}{Head-of-Line}
\newacronym{ia}{IA}{Initial Access}
\newacronym{imt}{IMT}{International Mobile Telecommunication}
\newacronym{iot}{IoT}{Internet of Things}
\newacronym{los}{LOS}{Line-of-Sight}
\newacronym{lte}{LTE}{Long Term Evolution}
\newacronym{m2m}{M2M}{Machine to Machine}
\newacronym{mac}{MAC}{Medium Access Control}
\newacronym{mc}{MC}{Multi-Connectivity}
\newacronym{mcs}{MCS}{Modulation and Coding Scheme}
\newacronym{mec}{MEC}{Mobile Edge Cloud}
\newacronym{mi}{MI}{Mutual Information}
\newacronym{mimo}{MIMO}{Multi-input multi-output}
\newacronym{mmwave}{mmWave}{millimeter wave}
\newacronym{mptcp}{MPTCP}{Multipath TCP}
\newacronym{mr}{MR}{Maximum Rate}
\newacronym{mss}{MSS}{Maximum Segment Size}
\newacronym{mtd}{MTD}{Machine-Type Device}
\newacronym{mtu}{MTU}{Maximum Transmission Unit}
\newacronym{nfv}{NFV}{Network Function Virtualization}
\newacronym{nlos}{NLOS}{Non-Line-of-Sight}
\newacronym{nr}{NR}{New Radio}
\newacronym{ofdm}{OFDM}{Orthogonal Frequency Division Multiplexing}
\newacronym{pdcch}{PDCCH}{Physical Downlink Control Channel}
\newacronym{pdcp}{PDCP}{Packet Data Convergence Protocol}
\newacronym{pdsch}{PDSCH}{Physical Downlink Shared Channel}
\newacronym{pdu}{PDU}{Packet Data Unit}
\newacronym{pf}{PF}{Proportional Fair}
\newacronym{pgw}{PGW}{Packet Gateway}
\newacronym{phy}{PHY}{Physical-layer}
\newacronym{pbch}{PBCH}{Physical Broadcast Channel}
\newacronym[plural=\gls{mme}s,firstplural=Mobility Management Entities (MMEs)]{mme}{MME}{Mobility Management Entity}
\newacronym{prb}{PRB}{Physical Resource Block}
\newacronym{pss}{PSS}{Primary Synchronization Signal}
\newacronym{pu}{PU}{primary users}
\newacronym{su}{SU}{secondary user}
\newacronym{pucch}{PUCCH}{Physical Uplink Control Channel}
\newacronym{pusch}{PUSCH}{Physical Uplink Shared Channel}
\newacronym{rach}{RACH}{Random Access Channel}
\newacronym{ran}{RAN}{Radio Access Network}
\newacronym{red}{RED}{Random Early Detection}
\newacronym{rf}{RF}{radio frequency}
\newacronym{rlc}{RLC}{Radio Link Control}
\newacronym{rlf}{RLF}{Radio Link Failure}
\newacronym{rrc}{RRC}{Radio Resource Control}
\newacronym{rrm}{RRM}{Radio Resource Management}
\newacronym{rr}{RR}{Round Robin}
\newacronym{rs}{RS}{Remote Server}
\newacronym{rsrp}{RSRP}{Reference Signal Received Power}
\newacronym{rss}{RSS}{Received Signal Strength}
\newacronym{rtt}{RTT}{Round Trip Time}
\newacronym{rw}{RW}{Receive Window}
\newacronym{rx}{RX}{Receiver}
\newacronym{sa}{SA}{standalone}
\newacronym{sack}{SACK}{Selective Acknowledgment}
\newacronym{sap}{SAP}{Service Access Point}
\newacronym{sch}{SCH}{Secondary Cell Handover}
\newacronym{scoot}{SCOOT}{Split Cycle Offset Optimization Technique}
\newacronym{fpga}{FPGA}{field-programmable gate array}
\newacronym{sdr}{SDR}{software-defined radio}
\newacronym{sdma}{SDMA}{Spatial Division Multiple Access}
\newacronym{sinr}{SINR}{Signal to Interference plus Noise Ratio}
\newacronym{sm}{SM}{Saturation Mode}
\newacronym{snr}{SNR}{Signal-to-Noise-Ratio}
\newacronym{son}{SON}{Self-Organizing Network}
\newacronym{ss}{SS}{Synchronization Signal}
\newacronym{srs}{SRS}{Sounding Reference Signal}
\newacronym{sss}{SSS}{Secondary Synchronization Signal}
\newacronym{tb}{TB}{Transport Block}
\newacronym{tcp}{TCP}{Transmission Control Protocol}
\newacronym{tdd}{TDD}{Time Division Duplexing}
\newacronym{tdma}{TDMA}{Time Division Multiple Access}
\newacronym{tfl}{TfL}{Transport for London}
\newacronym{tm}{TM}{Transparent Mode}
\newacronym{trp}{TRP}{Transmitter Receiver Pair}
\newacronym{tti}{TTI}{Transmission Time Interval}
\newacronym{ttt}{TTT}{Time-to-Trigger}
\newacronym{tx}{TX}{Transmitter}
\newacronym{ue}{UE}{User Equipment}
\newacronym{ul}{UL}{Uplink}
\newacronym{uml}{UML}{Unified Modeling Language}
\newacronym{um}{UM}{Unacknowledged Mode}
\newacronym{utc}{UTC}{Urban Traffic Control}
\newacronym{vm}{VM}{Virtual Machine}
\newacronym{rsrq}{RSRQ}{Reference Signal Received Quality}
\newacronym{rssi}{RSSI}{Received Signal Strength Indicator}
\newacronym{crs}{CRS}{Cell Reference Signal}
\newacronym{nsa}{NSA}{Non Stand Alone}
\newacronym{mrdc}{MR-DC}{Multi \gls{rat} \gls{dc}}
\newacronym{endc}{EN-DC}{E-UTRAN-\gls{nr} \gls{dc}}
\newacronym{5gc}{5GC}{5G Core}
\newacronym{si}{SI}{Study Item}
\newacronym{iab}{IAB}{Integrated Access and Backhaul}
\newacronym{wf}{WF}{Wired-first}
\newacronym{hqf}{HQF}{Highest-quality-first}
\newacronym{pa}{PA}{Position-aware}
\newacronym{mlr}{MLR}{Maximum-local-rate}
\newacronym{wbf}{WBF}{Wired Bias Function}
\newacronym{mib}{MIB}{Master Information Block}
\newacronym{sib}{SIB}{Secondary Information Block}
\newacronym{kpi}{KPI}{Key Performance Indicator}
\newacronym{ppp}{PPP}{Poisson Point Process}
\newacronym{gtp}{GTP}{GPRS Tunneling Protocol}
\newacronym{amf}{AMF}{Access and Mobility Management Function}
\newacronym{dash}{DASH}{Dynamic Adaptive Streaming over HTTP}
\newacronym{http}{HTTP}{HyperText Transfer Protocol}
\newacronym{qos}{QoS}{Quality of Service}
\newacronym{udp}{UDP}{User Datagram Protocol}
\newacronym{cu}{CU}{Central Unit}
\newacronym{du}{DU}{Distributed Unit}
\newacronym{mt}{MT}{Mobile Termination}
\newacronym{sdap}{SDAP}{Service Data Adaptation Protocol}
\newacronym{tdm}{TDM}{Time Division Multiplexing}
\newacronym{fdm}{FDM}{Frequency Division Multiplexing}
\newacronym{sdm}{SDM}{Space Division Multiplexing}
\newacronym{dag}{DAG}{Directed Acyclic Graph}
\newacronym{st}{ST}{Spanning Tree}
\newacronym{ummimo}{UM-MIMO}{Ultra-massive Multiple Input, Multiple Output}
\newacronym{wlan}{WLAN}{Wireless LAN}
\newacronym{wlans}{WLANs}{Wireless Local Area Networks}
\newacronym{rlnc}{RLNC}{Random Linear Network Coding}
\newacronym{drx}{DRX}{Discontinuous Reception}
\newacronym{cpu}{CPU}{Central Processing Unit}
\newacronym{soc}{SoC}{system-on-chip}
\newacronym{dcm}{DCM}{distributed cooperative \gls{mimo}}
\newacronym{comp}{CoMP}{Coordinated Multi-Point}
\newacronym{ap}{AP}{Access Point}

\newacronym{ssm}{SSM}{spectrum sensing module}
\newacronym{dsa}{DSA}{dynamic spectrum access}
\newacronym{bpm}{BPM}{baseband processing module}
\newacronym{drl}{DRL}{deep reinforcement learning}
\newacronym{5gb}{5GB}{5G and beyond}
\newacronym{ml}{ML}{machine learning}
\newacronym{cbrs}{CBRS}{Citizens Broadband Radio Service}
\newacronym{gaa}{GAA}{General Authorized Access}
\newacronym{pal}{PAL}{Priority Access Licensee}
\newacronym{fcc}{FCC}{Federal Communications Commission}
\newacronym{rfp}{RFP}{radio fingerprinting}
\newacronym{dsp}{DSP}{digital signal processing}
\newacronym{ssa}{SSA}{spectrum sensing and access}
\newacronym{fir}{FIR}{finite impulse response}
\newacronym{wsc}{WSC}{wireless signal classification}
\newacronym{ber}{BER}{bit error rate}
\newacronym{wwa}{WWA}{wideband waveform authentication}
\newacronym{ppi}{PPI}{protocol and parameter inference}

\newacronym{lna}{LNA}{low-noise amplifier}
\newacronym{lpf}{LPF}{low-pass filter}
\newacronym{bpf}{BPF}{band-pass filter}
\newacronym{rfic}{RFIC}{radio frequency integrated circuit}
\newacronym{adc}{ADC}{analog-to-digital converter}
\newacronym{agc}{AGC}{automatic gain control}
\newacronym{rrtml}{RadioRTML}{Radio Real-Time Machine Learning}
\newacronym{pcb}{PCB}{printed circuit board}
\newacronym{ser}{SER}{symbol error rate}
\newacronym{asic}{ASIC}{application-specific integrated circuit}
\newacronym{hdl}{HDL}{hardware description language}
\newacronym{hls}{HLS}{high-level synthesis}
\newacronym{pl}{PL}{programmable logic}
\newacronym{ps}{PS}{processing system}

\newacronym{evm}{EVM}{error vector magnitude}
\newacronym{mer}{MER}{modulation error ratio}
\newacronym{fec}{FEC}{forward error correction}
\newacronym{per}{PER}{packet error rate}

\newacronym{cmos}{CMOS}{complementary metal-oxide semiconductor}
\usepackage{amsmath,amsfonts}
\usepackage{algorithmic}
\usepackage{algorithm}
\usepackage{array}
\usepackage[caption=false,font=normalsize,labelfont=sf,textfont=sf]{subfig}
\usepackage{textcomp}
\usepackage{stfloats}
\usepackage{url}
\usepackage{verbatim}
\usepackage{graphicx}
\usepackage{cite}
\usepackage{makecell}
\usepackage{xcolor}
\usepackage{soul}

\newcommand{\mh}[1]{\ifthenelse{\boolean{showcomments}} {\begin{color}{orange}\textbf{#1}\end{color}} {}}

\begin{document}

\title{Toward Wireless System and Circuit Co-Design \\ for the Internet of Self-Adaptive Things}   

\author{\IEEEauthorblockN{Diptashree Das, Mohammad Abdi, Minghan Liu, Marvin Onabajo and Francesco Restuccia}\\
\vspace{-0.5cm}
 \IEEEauthorblockA{Department of Electrical and Computer Engineering, Northeastern University, United States}\vspace{-0.8cm}
 }
\thispagestyle{fancy}

\maketitle

\begin{abstract}  
The deployment of a growing number of devices in \gls{iot} networks implies that uninterrupted and seamless adaptation of wireless communication parameters (e.g., carrier frequency, bandwidth and modulation) will become essential. To utilize wireless devices capable of switching several communication parameters requires real-time self-optimizations at the \gls{rfic} level based on system level performance metrics during the processing of complex modulated signals. This article introduces a novel design verification approach for reconfigurable RFICs based on end-to-end wireless system-level performance metrics while operating in a  dynamically changing communication environment. In contrast to prior work, this framework includes two modules that simulate a wireless channel and decode waveforms. These are connected to circuit-level modules that capture device- and circuit-level non-idealities of RFICs for design validation and optimization, such as transistor noises, intermodulation/harmonic distortions, and memory effects from parasitic capacitances. We demonstrate this framework with a receiver (RX) consisting of a reconfigurable complementary metal-oxide semiconductor (CMOS) low-noise amplifier (LNA) designed at the transistor level, a behavioral model of a mixer, and an ideal filter model. The seamless integration between system-level wireless models with circuit-level and behavioral models (such as VerilogA-based models) for RFIC blocks enables to preemptively evaluate circuit and system designs, and to optimize for different communication scenarios with adaptive circuits having extensive tuning ranges. An exemplary case study is presented, in which simulation results reveal that the LNA power consumption can be reduced up to 16x depending on system-level requirements. \smallskip 
\end{abstract}

 \begin{IEEEkeywords}
 System-level validation, adaptive wireless systems and circuits, hardware/software co-design, energy-aware optimization, simulation-based system testing.
  \end{IEEEkeywords}

\begin{figure*}[t]
    \centering
    \includegraphics[width=1.94\columnwidth]{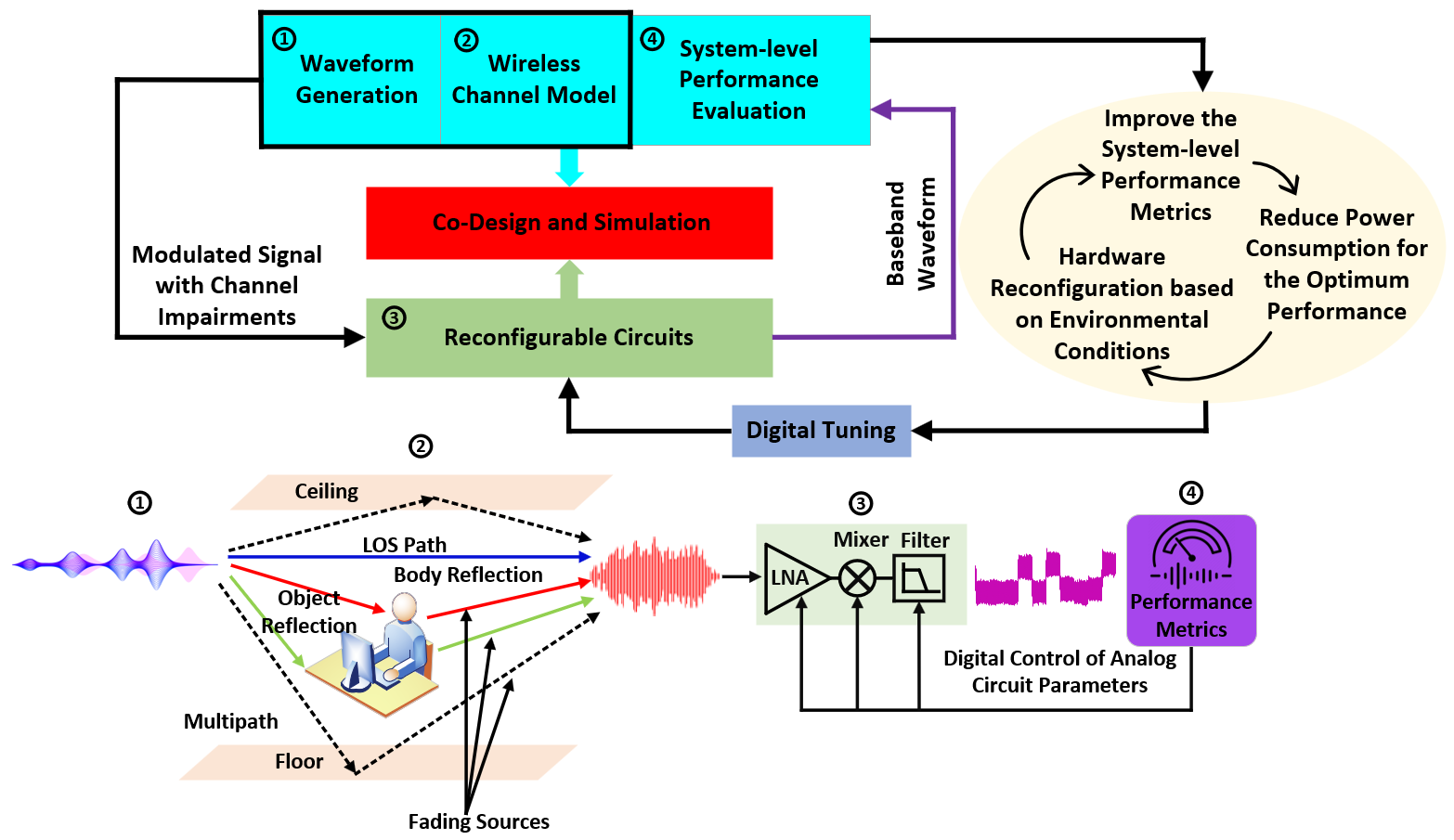}
    \setlength\abovecaptionskip{-.04cm}
    \caption{Overview of the framework for co-design and optimization of reconfigurable RF circuits and wireless communication systems.\vspace{-1cm}}
    \label{fig.structure}
\end{figure*}

\glsresetall

\section{Introduction}
\thispagestyle{fancy}

 \IEEEPARstart{B}y 2030, over 50 billion devices will be absorbed into the \gls{iot} \cite{SpectrumCrunch}. The sheer number of \gls{iot} devices implies that continuous and seamless adaptation of wireless communication parameters (e.g., carrier frequency, bandwidth and modulation) will become essential. One of the wireless system design goals is to provide sensing services for a plethora of applications. However, the implementations are constrained because sensing and communication circuit parameters have to be optimized for different frequency bands, modulation schemes and channel conditions \cite{LinSensing}. For this reason, \textit{reconfigurability} and \textit{self-optimization} will be a cornerstone of \gls{iot} networking paradigms \cite{Pena2017,chenDSS, justinDSS}. Furthermore, it becomes increasingly important to jointly simulate circuit and system level components for design optimization and validation prior to the fabrication of chips. On the other hand, conventional \glspl{rfic} are still statically optimized, which does not allow for real-time self-optimization at the intersection of hardware and software. Low-power \glspl{rfic} are typically designed and optimized specifically for the worst-case scenario of a given communication standard, which leads to performance limitations and excessive power consumption. Furthermore, circuit-level tuning usually optimizes block-level performance. Conversely, in real systems, the circuit-level linearity and dynamic range requirements strongly depend on the presence of nearby interference signals and bias conditions \cite{m45,murphy2012,andrew2010,Kim2008, RuJSSC}, while slow-varying aspects such as temperature sensitivity or device-level aging effects can only be computed and compensated throughout the device lifetime \cite{reddy2015aging}. Digitally-controlled calibration is a popular design approach to improve the performance and testability of mixed-signal integrated circuits \cite{Chauhan2016DandT, Monabajo2012}. However, RFIC calibrations of multiple interconnected circuit blocks typically do not address the interdependence of the circuit-specific parameters during tuning, which creates limitations during simulations for design validations. They normally also do not account for system-level parameters such as \gls{ser} or throughput, particularly when relying on single-tone/two-tone signals or other alternative test signals instead of the actual modulated signals \cite{m17}. A comprehensive survey about the integration of \gls{ml} into integrated circuit design has been provided in \cite{chen2013DandT,rapp2022}. The use of \gls{ml} during design and optimization can be another potential method to ensure functionality under consideration of interdependence between circuit and system parameters. 

Considering the above-mentioned challenges and opportunities, we propose a new RFIC co-design and validation paradigm at the intersection of hardware and software, which is summarized in Fig. 1. Our joint simulation framework considers \textit{system-level performance metrics} to facilitate the design of adaptive \gls{rfic} circuits. The developed interoperability between different design tools allows to (i) generate arbitrary modulated waveforms while modeling different wireless channel conditions, (ii) utilize foundry-supplied models that capture transistor-level non-idealities, (iii) simulate both behavioral analog RFIC blocks and circuit level designs, and (iv) extract circuit-level and system-level performance metrics. Furthermore, we introduce simulations with modulated signal packets and modeled channel impairments to extract system-level parameters such as  \gls{ber} and \gls{evm}; and accordingly optimize several circuit conditions (e.g., gain, noise figure and linearity characteristics) for optimization of energy/performance tradeoffs. As a use-case scenario, we leverage our framework to optimize a receiver (RX) composed of a reconfigurable \gls{cmos} \gls{lna} designed at the transistor-level, a behavioral model of a mixer and an ideal filter model. The simulation results show that our framework can reduce the \gls{lna} power consumption by up to 16x under varying \gls{ber} requirements. In general, the simulation framework can be used as a tool during the design and validation of adaptive wireless RXs that operate with dynamically changing requirements. \vspace{-0.1cm} 

\section{Existing Work and Current Challenges}

The fast-changing IoT ecosystem leads to a very dynamic nature of the wireless channel that calls for complex hardware and software systems, including adaptive and tunable transceiver designs. It has been explained in \cite{Leo2019Harware} how tunable and reconfigurable radio frequency (RF) technologies provide potential solutions for efficient spectrum sharing. In \cite{Luisopto}, a simulation-based approach develops a multi-user IoT communication system by taking into account carrier synchronization and data broadcasts on multiple channels for several of low‐power devices present in the network. The work in \cite{SoltaniDy} realized a learning-based RF signal classifier on a field-programmable gate array (FPGA) to reduce latency and power consumption, which requires prior knowledge of signals and spectrum. 

Adaptability in the RX front-end opens up the opportunity to collect and process data from a dynamic wireless channel. CMOS RFIC prototypes have been designed to enhance linearity and power handling requirements for cellular applications \cite{LiemMTT}. However, RFICs can still be complemented with real-time adaptation algorithms to optimize transceiver operation. \cite{MegTcas1} introduces a real-time two-dimensional real-time adaptation method to configure a RX for optimum NF and linearity with a certain power budget and desired signal level. The design of adaptive wireless RXs with single or multi-parameter optimization is highly relevant for specific incoming signals, and requires validating performance for different wireless standards, channel conditions, and chip-level performance variations (e.g., CMOS fabrication process variations). A key consideration during the design of adaptive RFICs is that reconfigurability is tightly coupled with power consumption. In addition, the need for wide tuning range and energy/power scalable designs calls for the seamless combination of circuit and system level adjustments. An evaluation of performance versus power trajectory for RF front-end functional blocks has been explored in \cite{changEmerg}. In addition to determining the inter-dependencies of circuit parameters for each block in the RF front-end, the optimization of analog RF circuits based on feedback control with digitally-controlled features has been demonstrated \cite{DevVLSI, FengJssc,Murmann,LiVLSi,BanerjeeJssc, Monabajo2012}, which requires to design complex control strategies for different conditions in the presence of channel and device level variations. 

The problem of energy efficiency and channel conditions is conventionally controlled by adaptive modulation and coding \cite{Goldsmith}, making it harder for the RF front-end to adapt to any changes in the channel. Integrating tunable RFICs into spectrum-agile wireless networks can allow to self-optimize RF circuit parameters to produce a desired output signal within the optimum power budget based on existing channel conditions. \cite{SenTCAD} describes a channel-adaptive RX design with process variation tolerance. Furthermore, a neural network based self-learning RF system has been demonstrated in \cite{BanerjeeTCAS1}, which is able to reduce power consumption of wireless transceiver systems by dynamically tuning the circuit components while monitoring the effects of real-time wireless channel conditions and the fabrication process variations to produce a desired \gls{ber} and threshold \gls{evm}. This is achieved with an on-chip look-up table that requires to be updated based on expected channel conditions. 

The difficulty of simulating the entire system is a major impediment to verify the merits of the integrated hardware-software based wireless system. This work aims to design and validate a joint simulation platform that can address the inter-dependencies of circuit parameters in the RF front-end to develop self-optimized wide-range reconfigurable RX architecture together with wireless network that is capable of changing communication parameters. Furthermore, this approach to incorporate and verify system-level performance-driven tuning features using reconfigurable RFIC blocks is especially compelling to enhance resilience  to sudden changes in the environment and accordingly optimize to achieve performance targets with optimum power consumption. Hence, the presented simulation framework is expected to ease the adaptive design and optimization of closed-loop self-supervised Internet of Self-adaptive Things with different modulated signals, wireless channel models and adaptive RFIC designs together with circuit level non-idealities.

\section{Optimization Framework Overview}

We consider a scenario as depicted in Fig. 1, where we model multipath effects and dynamic fading along with arbitrary modulated signal packets during the design of robust RFIC adaptability to receive and process information. Here, the waveform generation allows to change the modeled \gls{phy} parameters (such as modulation scheme, power carried by the spectrum components, RF sampling frequency, channel coding and bandwidth). Currently, only  the \gls{snr} is used as a primary indicator of variable channel conditions. The variable SNR-based model encompasses several channel impairments such as additive white Gaussian noise (AWGN), multi-path fading, variable distance between transmitter (TX) and RX, and path-loss among other interference as indicated in Fig. 1. The sensitivity requirement of the RF front-end strongly depends on the SNR. \cite{Shi2019DL} includes a  description of typical SNR values for channels, where the comparative results with different modulation schemes provide insights into the expected SNR values for various channel conditions. It is also envisioned that future spectrum-agile TXs will lead to more variations of interference levels, SNR values, and modulation schemes.


As depicted in Fig. 1, the baseband waveform is processed to infer parameters associated with system-level performance such as \gls{ber}, \gls{ser}, EVM, \gls{mer} and \gls{per}. This approach is based on the goal to develop algorithms for accurate data-driven optimization of RFICs based on system performance. Next, we present an architecture with a reconfigurable RF front-end circuit that is evaluated through the simulation framework from this work. 

\begin{figure*}[t]
    \centering
    \includegraphics[width=2\columnwidth]{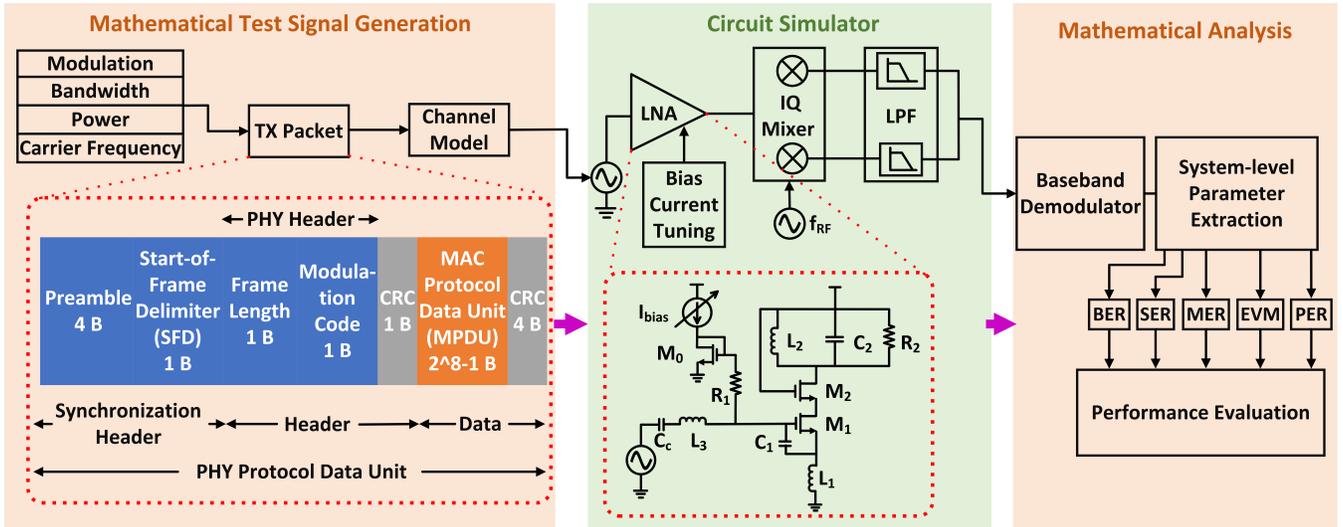}
    \setlength\abovecaptionskip{-.04cm}
    \caption{Co-design simulation platform for wireless systems with dynamically reconfigurable RF front-end circuits.\vspace{-1cm}}
\end{figure*}

\section{Design and Modeling of RF Front-end Circuits with Dynamic Reconfigurability}

\subsection{Flexible RF Front-end Architecture}

Analog RF front-end reconfigurability enhances co-existence and spectrum sharing in crowded environments \cite{MohamedWireless}, as well as allows to vary parameters such as data rates on demand \cite{MahmoudTcas1}. As shown in Fig. 2, the reconfigurable RF RX front-end in this work consists of a digitally programmable LNA circuit designed at the transistor-level with tunable bias current, a behavioral model of a direct down-conversion in-phase/quadrature (I/Q) mixer stage, and two ideal low-pass filter (LPF) models in the I and Q paths. Circuit-level simulations with device-level non-idealities can capture impacts of frequency response limitations, inter-modulation products, thermal and flicker noises, parasitic capacitances/resistances, and higher-order non-linearities of the transistors, that allows  block-level specifications assessment such as gain, noise figure (NF), input third-order inter-modulation intercept point (IIP3) and impedance matching conditions. At the same time, the ability to include some behavioral models of circuit blocks aids the early design and system-level verification phase. Most importantly,  the accurate transient simulations allow to account for the impacts of circuit-level imperfections to assess system-level metrics with changing environment.

 To overcome existing inflexible wireless standards, inefficient spectrum use and potential security threats in the wireless network, the flexible adaptation of \gls{phy} parameters of the signal proves an effective and long-standing solution. The work in \cite{Vo2016Fingerprinting} demonstrated if TXs were allowed to dynamically switch \gls{phy} parameters such as carrier frequency and symbol modulation, the TXs would become less jamming-prone and achieve more efficient spectrum occupation. To give an example, Fig. 2 shows the selected test signal with a format that corresponds to the Zigbee \gls{phy} packet structure. This signal with any modeled channel impairments has been applied as input signal during circuit and behavioral simulations of the RF front-end. The frame starts with a known preamble for synchronization, which exhibits high auto-correlation and low cross-correlation features. The preamble is followed by a start-of-frame delimiter that marks the beginning of the header. The header consists of the frame length in bytes and the modulation code associated with the modulation scheme used. A data checksum (CSC) is attached to the header and data parts respectively, such that erroneous frames can be detected and discarded. The data part of the frame can support a MAC Protocol Data Unit (MPDU) with a size of up to $2^8-1$ bytes.

As depicted in Fig. 2, the emulated transmitted packets with added channel noise and imperfections are transferred to the signal source of a circuit simulator (Cadence Spectre) for the RF front-end simulation. The model-based design simulator (Matlab) is capable of saving the raw data in the comma-separated values (CVS) file format, which has been incorporated into the circuit simulator by the virtue of a piece-wise linear signal source. Thus, the reconfigurability settings and biasing conditions of the RF front-end blocks can be evaluated during the circuit design phase to optimize characteristics such as noise levels, linearity, RX sensitivity and impedance matching conditions. As mentioned earlier, the mathematical analysis of the circuit simulator output provides system-level performance metrics (such as \gls{ber}, \gls{ser}, \gls{mer}, and \gls{evm}) for both circuit and system-level optimization with modulated signals. Here, the \gls{ber} is considered as one of the most significant \gls{phy} performance indicators. Since not all inaccuracies lead to bit flips, \gls{evm} is another appropriate measure to quantify the quality of the received signal after processing in the RF front-end, which allows to capture important channel and RX non-idealities \cite{AcarTCADEVM, NatarajanEVM}. Various imperfections such as changing channel conditions have impacts on the \gls{evm} since they can cause the received constellation points to deviate from their original ideal locations. 

\subsection{Reconfigurable LNA}

A 2.4 GHz single-ended cascode common-source LNA with inductive source degeneration has been selected as an exemplary reconfigurable narrowband LNA design, of which the bias current is tuned to control performance/power tradeoffs as shown in Fig. 2. Since the LNA is the first block of the RF front-end, its performance is particularly crucial when receiving noisy packets at low power levels. The standalone LNA was designed in a standard 65nm CMOS technology, and simulated for bias currents ranging from 31.25 $\mu$A to 500 $\mu$A. By changing the bias current +/-20\% to +/-50\% from its design point (125 $\mu$A), the circuit characteristics such as gain, IIP3, and NF can be adjusted with a corresponding change of the power consumption that is directly proportional to the bias current. Fig. 3 summarizes the gain, NF and IIP3 of the LNA from transistor-level  simulations. 

\begin{figure}[h]
    \centering
    \includegraphics[width=1\columnwidth]{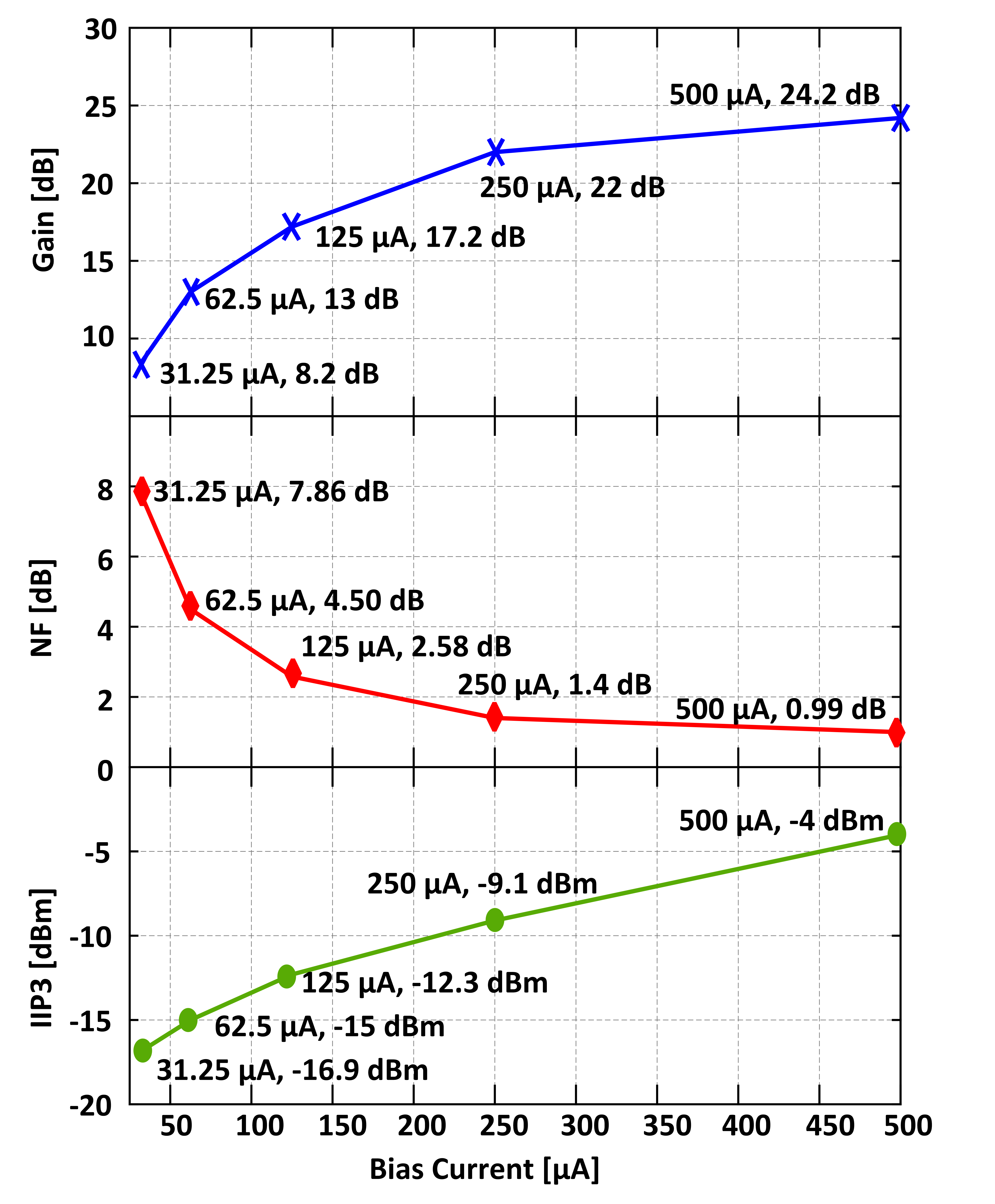}
    \setlength\abovecaptionskip{-0.5cm}
    \caption{Simulated LNA performance parameters vs. bias current.\vspace{-0.5cm}}
\end{figure}

\subsection{Mixer and Baseband Signal Processing}

To portray the capability of combining transistor-level schematic simulations (i.e., the LNA) with behavioral blocks during transient circuit simulations to assess system-level metrics under specified/changeable wireless channel conditions, a direct down-conversion I/Q mixer has been modeled in VerilogA, followed by two ideal LPFs (one in each RX path). The mixer model includes variable gain, IIP3 and NF based on typical reconfigurable circuit parameters of down-conversion mixers. In this proof-of-concept, the performance assessment of the RF front-end was with QPSK-modulated signals and channel impairments according to the signal generation in Fig. 2. 

The channel model accounts for AWGN, path-loss and generic frequency-selective multi-path fading that can introduce different path attenuation, delay, and Doppler shift. The baseband demodulator after the LPF is also implemented mathematically, where according to the decision boundaries defined by the associated constellation, the I/Q samples are detected and compared against the ground-truth I/Q samples to compute the desired system-level parameters. Future work will be devoted to automatic parameter tuning parameters in the RF front-end circuits based on the extracted system-level performance metrics to optimize with varying spectrum conditions under specified power consumption targets. 

\section{Case Study: \\ Reconfigurable RF Front-end Simulation}

\begin{figure*}[t]
    \centering
    \includegraphics[width=2\columnwidth]{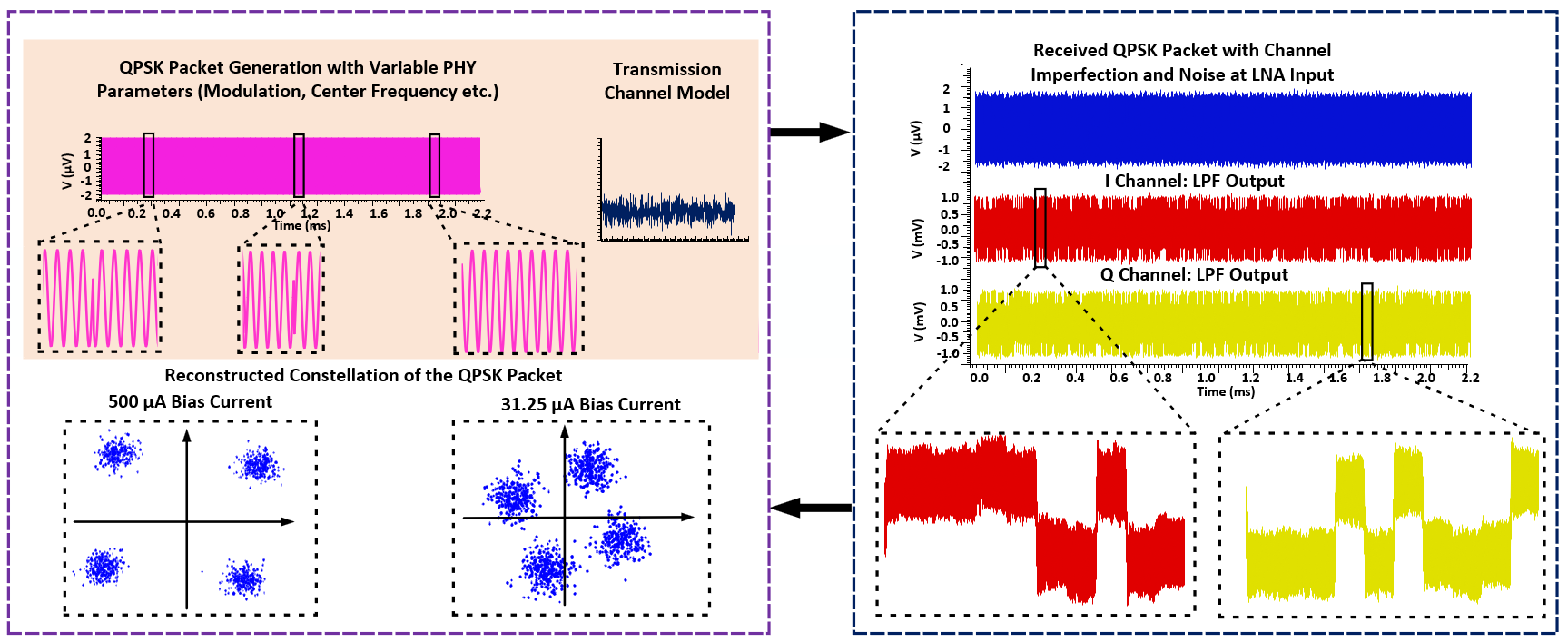}
    \setlength\abovecaptionskip{-.04cm}
    \caption{Results from the simulation with generated QPSK packets: transient RF front-end characteristics and reconstructed constellations.\vspace{-0.01cm}}
\end{figure*}

\begin{table*}[t]
    \centering
    \caption{System-level performance summary for the re-configurable RF front-end with LNA bias tuning from simulations with a QPSK burst (4272 bits) having a power of -100 dBm.\vspace{-0.05cm}}
    \includegraphics[width=1.94\columnwidth]{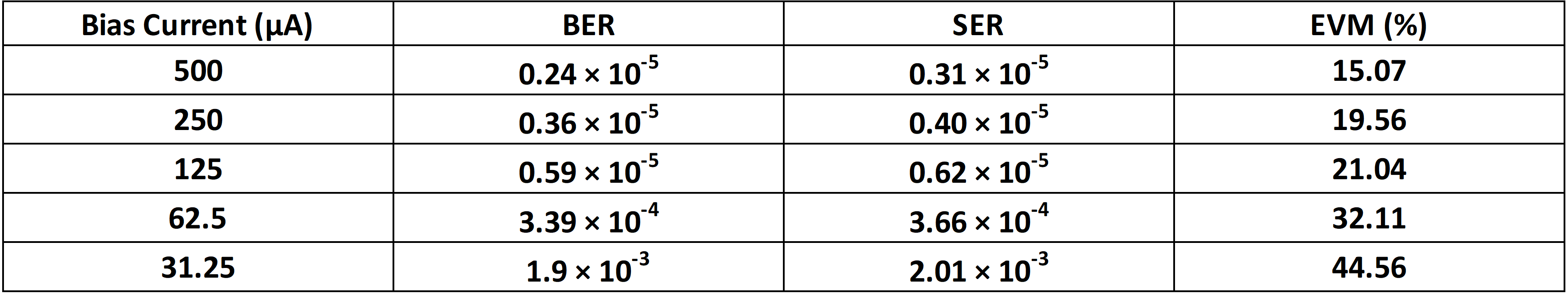}
    \setlength\abovecaptionskip{-.04cm}    
\end{table*}

This section summarizes results from the use of our co-simulation framework for the example RF front-end configuration described in the previous section. The simulations were primarily carried out to evaluate the performance vs. power trade-offs associated with the reconfigurable LNA design. As mentioned in the previous section, the complete testbench used for the simulations includes a VerilogA based mixer and ideal LPF. The mixer has been modeled with flexible circuit parameters in which the gain, IIP3 and NF can be changed as part of the design exploration. In this work, we have used an ideal behavioral mixer model with a gain of 10 dB, IIP3 of 5 dBm and NF of 10dB for the proof-of-concept simulations \cite{KasTcasII, LehneISCAS, HaoTCASI}. The upper left image in Fig. 4 displays the generated QPSK-modulated signal with the packet structure defined in Fig. 2. A binary ground-truth message is randomly generated and up-converted to produce the QPSK modulation with a center frequency of 2.4 GHz. The SNR of the received signal was selected as 20 dB to emulate typical wireless network conditions with channel impairments and distortion \cite{Garcıa-Naya2011, Zhang2014}. The corresponding waveform of the generated QPSK signal in Fig. 4 is corrupted by the channel imperfections and noise, and fed to the RF front-end. The simulated I/Q signals at the LPF outputs are shown on the right side of Fig. 4. These I/Q signals are transferred for mathematical baseband processing using an envelope detector and moving average filter (as displayed in Fig. 2) during model-based simulations.

The demodulated I/Q samples are then compared and verified with the ground-truth data to extract \gls{ber}, \gls{ser}, \gls{evm} and \gls{mer}. The lower values of the \gls{ber} and \gls{ser} relust from high accuracy of demodulation process. In this case, the BER and SER are in the range of $10^{-5}$-$10^{-3}$ for all LNA bias current conditions. On the other hand, the MER values are 18.9 dB and 11.2 dB for LNA bias currents of 500 $\mu$A and 31.25 $\mu$A respectively. The proximity of the MER value to the specified channel SNR (20 dB) is an indication of a noise resilient system, which results from the LNA bias with high current (i.e., high power consumption) to achieve a low NF. Fig. 4 includes the extracted constellations of the QPSK signal after the processing by the RF front-end with the highest and lowest LNA bias currents, which also show the power vs. performance tradeoff. Table I includes an overview of the system-level performance for different LNA bias current settings. We have simulated two QPSK packets (4272 bits) with randomly generated ground-truth data to evaluate the high-level system performance with the co-design platform. It can be seen from Table I that the BER, SER and EVM are considerably lower for LNA bias currents in the 125 $\mu$A to 500 $\mu$A range. In addition, no packet errors (PEs) occurred in the 125 $\mu$A to 500 $\mu$A bias current range. The simulation results show robust adaptability, which will be realized with an application-specific feedback control loop as depicted in Fig. 1. Depending on the application-specific \gls{ber} requirement and channel conditions, energy consumption can be significantly reduced through the 62.5 $\mu$A and 31.25 $\mu$A bias current settings. 

\section{Next Steps: Beyond Traditional Wireless System and RFIC Integration}

We foresee that the research presented in this paper will be the foundation for the development of reconfigurable RXs with real-time self-optimization capabilities. The work described in this article is the first step associated with the co-design and verification of wireless systems and circuits that can tolerate and adapt to interference conditions using novel RFIC optimization methods with unprecedented design flexibility based on specified system-level performance metrics. As depicted in Fig. 5, we anticipate that the presented co-simulation and design verification framework will contribute to the development of several novel features: (i) \gls{ml}-based self-decisive CMOS RF front-ends to adapt changing network conditions, (ii) development of hardware-software prototypes based on joint integrated circuit simulations and wireless data collection for enhanced modeling, (iii) several digitally-controlled tuning knobs in each analog block as indicated in Fig. 1, and (iv) collection of waveform datasets through the experimental testbenches to train ML algorithms. (v) Once the \gls{ml} algorithms are developed, the \gls{drl} agent will be trained to collect data, both experimentally and synthetically utilizing the proposed framework. During the data collection, one can deploy the optimal policy on FPGA-based platforms such as \glspl{sdr} \cite{restuccia2020deepwierl,XuCloudEEG} to meet the challenging time constraints involved, and to reduce the overall power consumption. The presented simulation framework features tools to jointly optimize the power-efficiency of digitally-controlled analog circuits and the computation resources to implement adaptive ML-based control. Once the envisioned \gls{ml} algorithm is developed and the \gls{drl} agent is trained experimentally and synthetically collected data using the proposed framework, one could deploy the optimal policy on FPGA-based platforms such as \glspl{sdr} \cite{restuccia2020deepwierl} so as to meet the challenging time constraints involved. To further minimize the power consumption introduced by running the \gls{ml} method, one can change the operational frequency or even limit it to the times when the system-level performance drops below a certain threshold or experiences a sudden change. The simulation results in Section V show that the RF front-end in this case study has a sensitivity of -100 dBm. We have presented a case study with a QPSK packet, but other modulation schemes (e.g. ASK, BPSK) with different SNR values, data rate, bandwidth, center frequencies can be employed. The reconfigurable RX design and simulation approach is intended to facilitate the use of flexible wireless system parameters by reconfiguring circuit parameters for the optimum performance and spectrum sharing in real-time.

\begin{figure}[h]
    \centering
    \includegraphics[width=1\columnwidth]{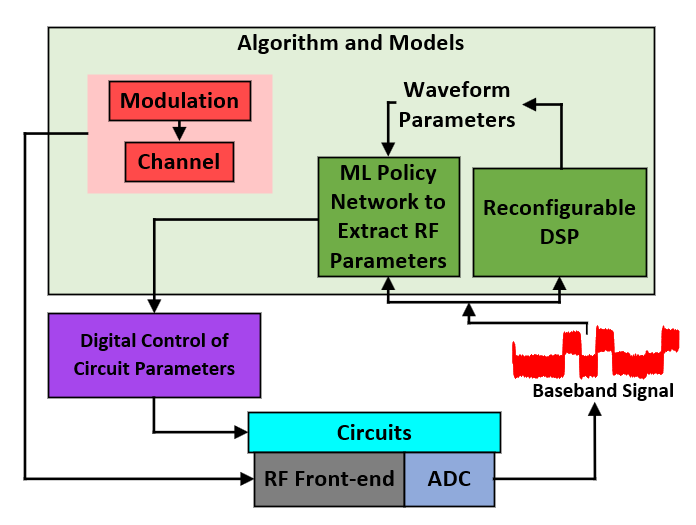}
    \setlength\abovecaptionskip{-0.4cm}
    \caption{Envisioned ML-based real-time RF front-end optimization.\vspace{-0.2cm}}
\end{figure}

                

\section{Conclusion}

This article presented a new co-design simulation platform to evaluate trade-offs between performance and power consumption by integrating the design and modeling of wireless systems and reconfigurable RF front-end circuits together. A main contribution of the integrated framework is to provide an effective and long-lasting tool for the design of spectrum-agile RXs with adaptive RFICs for dynamic optimizations under changing environmental conditions. The execution of the joint summation taking into account combining dynamic wireless channel model and reconfigurable RF front-end with circuit level non-idealities accounts for incorporation of several modeling and design software to validate the performance. The reconfigurability of the RF front-end aims to reduce power consumption significantly based on application-specific system-level performance targets. However, the primary goal is to execute the seamless adaptation of the RF front-end by optimizing its circuit parameters during real-time execution with spectrum-agile transmission to produce desired end-to-end system level performance.  The integration of digital tuning capabilities in each of the analog blocks within RFICs will be particularly useful for realizing future wireless system paradigms with an unprecedented degree-of-freedom.
\vspace{-0.1cm}

\section{Acknowledgement}

This material is in part based upon work supported by the NSF under grant no. ECCS-2146754, CCF-2218845, CNS-2134973, CNS-2120447, ECCS-2229472, and is supported in part by funds from OUSD R\&E, NIST, and industry partners as specified in the Resilient \& Intelligent NextG Systems (RINGS) program, as well as by the Air Force Office of Scientific Research under contract number FA9550-23-1-0261, and by the Office of Naval Research under award number N00014-23-1-2221. The U.S. Government is  authorized to reproduce and distribute reprints for Governmental purposes notwithstanding any copyright  notation thereon. The views and conclusions contained herein are those of the authors and should not be interpreted as necessarily representing the official policies or endorsements of the NSF and the U.S. Government.  

\bibliographystyle{IEEEtran}
\footnotesize
\bibliography{bibliography}

\begin{thebibliography}{10}
\providecommand{\url}[1]{#1}
\csname url@samestyle\endcsname
\providecommand{\newblock}{\relax}
\providecommand{\bibinfo}[2]{#2}
\providecommand{\BIBentrySTDinterwordspacing}{\spaceskip=0pt\relax}
\providecommand{\BIBentryALTinterwordstretchfactor}{4}
\providecommand{\BIBentryALTinterwordspacing}{\spaceskip=\fontdimen2\font plus
\BIBentryALTinterwordstretchfactor\fontdimen3\font minus \fontdimen4\font\relax}
\providecommand{\BIBforeignlanguage}[2]{{%
\expandafter\ifx\csname l@#1\endcsname\relax
\typeout{** WARNING: IEEEtran.bst: No hyphenation pattern has been}%
\typeout{** loaded for the language `#1'. Using the pattern for}%
\typeout{** the default language instead.}%
\else
\language=\csname l@#1\endcsname
\fi
#2}}
\providecommand{\BIBdecl}{\relax}
\BIBdecl

\bibitem{SpectrumCrunch}
{Federal Communications Commission (FCC)}, ``{Spectrum Crunch},'' \url{https://www.fcc.gov/general/spectrum-crunch}.

\bibitem{LinSensing}
A.~Liu, Z.~Huang, M.~Li, Y.~Wan, W.~Li, T.~Xiao-Han, C.~Liu, R.~Du, D.~Kai Pin~Tan, J.~Lu, Y.~Shen, F.~Colone, and K.~Chetty, ``A survey on fundamental limits of integrated sensing and communication,'' \emph{IEEE Communication Surveys and Tutorials}, vol.~24, no.~2, pp. 994 -- 1034, 2022.

\bibitem{Pena2017}
M.~D.~V. Pena, J.~J. Rodriguez-Andina, and M.~Manic, ``The internet of things: The role of reconfigurable platforms,'' \emph{IEEE Industrial Electronics Magazine}, vol.~11, no.~3, pp. 6--19, 2017.

\bibitem{chenDSS}
M.~Chen, A.~Liu, W.~Liu, K.~Ota, M.~Dong, and N.~N. Xiong, ``{RDRL: A Recurrent Deep Reinforcement Learning Scheme for Dynamic Spectrum Access in Reconfigurable Wireless Networks},'' \emph{IEEE Transactions on Network Science and Engineering}, vol.~9, no.~2, pp. 364 -- 376, 2022.

\bibitem{justinDSS}
J.~Tallon, T.~K. Forde, and L.~Doyle, ``{Dynamic Spectrum Access Networks: Independent Coalition Formation},'' \emph{IEEE Vehicular Technology Magazine}, vol.~7, no.~2, pp. 69 -- 76, 2012.

\bibitem{m45}
M.~{Abdulaziz}, E.~A.~M. {Klumperink}, B.~{Nauta}, and H.~{Sjöland}, ``Improving receiver close-in blocker tolerance by baseband {$G_m-C$} notch filtering,'' \emph{IEEE Transactions on Circuits and Systems I: Regular Papers}, vol.~66, no.~3, pp. 885--896, 2019.

\bibitem{murphy2012}
D.~Murphy, H.~Darabi, A.~Abidi, A.~A. Hafez, A.~Mirzaei, M.~Mikhemar, and M.-C.~F. Chang, ``A blocker-tolerant, noise-cancelling receiver suitable for wideband wireless applications,'' \emph{IEEE Journal of Solid-State Circuits}, vol.~47, no.~12, pp. 2943 -- 2963, 2012.

\bibitem{andrew2010}
C.~Andrews and A.~C. Molnar, ``A passive mixer-first receiver with digitally controlled and widely tunable rf interface,'' \emph{IEEE Journal of Solid-State Circuits}, vol.~45, no.~12, pp. 2696 -- 2708, 2010.

\bibitem{Kim2008}
N.~Kim, L.~E. Larson, and V.~Aparin, ``A highly linear saw-less cmos receiver using a mixer with embedded tx filtering for cdma,'' \emph{IEEE Transactions on Computer-Aided Design of Integrated Circuits and Systems}, 2008.

\bibitem{RuJSSC}
Z.~Ru, N.~A. Moseley, E.~A.~M. Klumperink, and B.~Nauta, ``Digitally enhanced software-defined radio receiver robust to out-of-band interference,'' \emph{IEEE Transactions on Computer-Aided Design of Integrated Circuits and Systems}, vol.~44, no.~12, pp. 3359--3375, 2009.

\bibitem{reddy2015aging}
S.~Arasu, M.~Nourani, J.~Carulli, and V.~Reddy, ``{Controlling Aging in Timing-Critical Paths},'' \emph{IEEE Design $\&$ Test}, vol.~33, no.~4, pp. 82--91, 2015.

\bibitem{Chauhan2016DandT}
H.~Chauhan, M.~Onabajo, V.~Kvartenko, R.~Coxe, and T.~Weber, ``An optimization platform for digital predistortion of power amplifiers,'' \emph{IEEE Design $\&$ Test}, vol.~33, no.~2, pp. 49--58, 2016.

\bibitem{Monabajo2012}
M.~Onabajo and J.~Silva-Martinez, ``{Analog Circuit Design for Process Variation-Resilient Systems-on-a-Chip},'' \emph{Springer}, 2012.

\bibitem{m17}
B.~Georgescu, R.~Salmeh, M.~Fattouche, and F.~Ghannouchi, ``Two-tone phase delay control of center frequency and bandwidth in low-noise-amplifier {RF} front ends,'' \emph{IEEE Transactions on Circuits and Systems II: Express Briefs}, vol.~60, no.~4, pp. 192--196, 2013.

\bibitem{chen2013DandT}
T.~Chen, G.~Zhang, B.~Yu, B.~Li, and U.~Schlichtmann, ``{Machine Learning in Advanced IC Design: A Methodological Survey},'' \emph{IEEE Design $\&$ Test}, vol.~40, no.~1, pp. 17--33, 2023.

\bibitem{rapp2022}
M.~Rapp, H.~Amrouch, Y.~Lin, B.~Yu, D.~Z. Pan, W.~Marilyn, and J.~Henkel, ``Mlcad: A survey of research in machine learning for cad keynote paper,'' \emph{IEEE Transactions on Computer-Aided Design of Integrated Circuits and Systems}, vol.~41, no.~10, pp. 3162 -- 3181, 2022.

\bibitem{Leo2019Harware}
L.~Laughlin, F.~Boccardi, G.~Chris, E.~Arabi, K.~C. Balram, K.~A. Morris, and M.~A. Beach, ``Emerging hardware enablers for more efficient use of the spectrum,'' \emph{2019 IEEE International Symposium on Dynamic Spectrum Access Networks (DySPAN)}, 2019.

\bibitem{Luisopto}
L.~Rodrigues, M.~Figueiredo, L.~Nero~Alves, and Z.~Ghassemlooy, ``Carrier synchronisation in multiband carrierless amplitude and phase modulation for visible light communication-based iot systems,'' \emph{IET Optoelectronics}, 2023.

\bibitem{SoltaniDy}
S.~Soltani, Y.~E. Sagduyu, R.~Hasan, K.~Davaslioglu, H.~Deng, and T.~Erpek, ``Real-time experimentation of deep learning-based rf signal classifier on fpga,'' \emph{2019 IEEE International Symposium on Dynamic Spectrum Access Networks (DySPAN)}, 2023.

\bibitem{LiemMTT}
B.-V. Liempd, B.~Hershberg, S.~Ariumi, K.~Raczkowski, K.-F. Bink, U.~Karthaus, E.~Martens, P.~Wambacq, and J.~Craninckx, ``A +70-dbm iip3 electrical-balance duplexer for highly integrated tunable front-ends,'' \emph{IEEE Transaction on Microwave Theory and Techniques}, vol.~64, no.~12, 2016.

\bibitem{MegTcas1}
M.~Meghdadi and M.~S. Bakhtiar, ``Two-dimensional multi-parameter adaptation of noise, linearity, and power consumption in wireless receivers,'' \emph{IEEE Transactions on Circuits and Systems I: Regular Papers}, vol.~61, no.~8, 2014.

\bibitem{changEmerg}
G.~Chang, S.~Maity, B.~Chatterjee, and S.~Sen, ``A medradio receiver front-end with wide energy-quality scalability through circuit and architecture-level reconfigurations,'' \emph{IEEE Journal on Emerging and Selected Topics in Circuits and Systems}, vol.~8, no.~3, 2018.

\bibitem{DevVLSI}
S.~Devarakond, S.~Sen, A.~Banerjee, and A.~Chatterjee, ``Digitally assisted built-in tuning using hamming distance proportional signatures in rf circuits,'' \emph{IEEE Transactions on Very Large Scale Integration (VLSI) Systems}, vol.~24, no.~9, p. 2918–2931, 2016.

\bibitem{FengJssc}
Y.~Feng, G.~Takemura, S.~Kawaguchi, N.~Itoh, and P.~R. Kinget, ``Digitally assisted iip2 calibration for cmos direct-conversion receivers,'' \emph{IEEE Journal of Solid-state Circuits}, vol.~46, no.~10, p. 2253–2267, 2011.

\bibitem{Murmann}
B.~Murmann, ``Digitally assisted analog circuits,'' \emph{IEEE Micro}, vol.~26, no.~2, pp. 38--47, 2006.

\bibitem{LiVLSi}
S.~Li, J.~Li, X.~Gu, H.~Wang, C.~Li, J.~Wu, and M.~Tang, ``Reconfigurable all-band rf cmos transceiver for gps/glonass/galileo/beidou with digitally assisted calibration,'' \emph{IEEE Transactions on Very Large Scale Integration (VLSI) Systems}, vol.~23, no.~9, pp. 1814 -- 1827, 2014.

\bibitem{BanerjeeJssc}
G.~Banerjee, M.~Behera, M.~A. Zeidan, R.~Chen, and K.~Barnett, ``Analog/rf built-in-self-test subsystem for a mobile broadcast video receiver in 65-nm cmos,'' \emph{IEEE Journal of Solid-State Circuits}, vol.~46, no.~9, pp. 1998 -- 2008, 2011.

\bibitem{Goldsmith}
A.~Goldsmith, ``Adaptive modulation and coding for fading channels,'' \emph{Proceedings of the 1999 IEEE Information Theory and Communications Workshop}, p. 24–26, 1999.

\bibitem{SenTCAD}
S.~Sen, V.~Natarajan, S.~Devarakond, and A.~Chatterjee, ``Process-variation tolerant channel-adaptive virtually zero-margin low-power wireless receiver systems,'' \emph{IEEE Transactions on Computer-Aided Design of Integrated Circuits and Systems}, vol.~33, no.~12, pp. 1764 -- 1777, 2014.

\bibitem{BanerjeeTCAS1}
D.~Banerjee, B.~Muldrey, , X.~Wang, S.~Sen, and A.~Chatterjee, ``Self-learning rf receiver systems: Process aware real-time adaptation to channel conditions for low power operation,'' \emph{IEEE Transactions on Circuits and Systems I: Regular Papers}, vol.~64, no.~1, pp. 195 -- 207, 2017.

\bibitem{Shi2019DL}
Y.~Shi, K.~Davaslioglu, Y.~E. Sagduyu, W.~C. Headley, M.~Fowler, and G.~Green, ``Deep learning for {RF} signal classification in unknown and dynamic spectrum environments,'' in \emph{2019 IEEE International Symposium on Dynamic Spectrum Access Networks (DySPAN)}, 2019, pp. 1--10.

\bibitem{MohamedWireless}
M.~Mohamed, S.~Handagala, J.~Xu, M.~Leeser, and M.~Onabajo, ``Strategies and demonstration to support multiple wireless protocols with a single rf front-end,'' \emph{IEEE Wireless Communications}, vol.~27, no.~3, pp. 88--95, 2020.

\bibitem{MahmoudTcas1}
M.~Ibrahim, Mahmoud A. A.and~Onabajo, ``A 0.061 nj/b 10 mbps hybrid bf-psk receiver for internet of things applications,'' \emph{IEEE Wireless Communications}, vol.~69, no.~5, pp. 1919 -- 1931, 2022.

\bibitem{Vo2016Fingerprinting}
T.~Vo-Huu, T.~Vo-Huu, and G.~Noubir, ``{Fingerprinting Wi-Fi Devices Using Software Defined Radios},'' \emph{in Proc. 9th ACM Conference on Security \& Privacy in Wireless and Mobile Networks}, pp. 3--14, 2016.

\bibitem{AcarTCADEVM}
E.~Acar and S.~Ozev, ``Low-cost characterization and calibration of rf integrated circuits through i–q data analysis,'' \emph{IEEE Transactions on Computer-Aided Design of Integrated Circuits and Systems}, vol.~28, no.~7, pp. 993 -- 1005, 2009.

\bibitem{NatarajanEVM}
V.~Natarajan, H.~W. Choi, A.~Banerjee, S.~Sen, A.~Chatterjee, G.~Srinivasan, F.~Taenzler, and S.~Bhattacharya, ``Low cost evm testing of wireless rf soc front-ends using multitones,'' \emph{IEEE Transactions on Computer-Aided Design of Integrated Circuits and Systems}, vol.~31, no.~7, pp. 1088 -- 1101, 2012.

\bibitem{KasTcasII}
M.~H. Kashani, M.~Asghari, M.~Yavari, and S.~Mirabbasi, ``A +7.6 dbm iip3 2.4-ghz double-balanced mixer with 10.5 db nf in 65-nm cmos,'' \emph{IEEE Transactions on Circuits and Systems II: Express Briefs}, vol.~68, no.~10, pp. 3214 -- 3218, 2021.

\bibitem{LehneISCAS}
M.~Lehne, J.~Stonick, and U.~Moon, ``An adaptive offset cancellation mixer for direct conversion receivers in 2.4 ghz cmos,'' \emph{2000 IEEE International Symposium on Circuits and Systems (ISCAS)}, 2000.

\bibitem{HaoTCASI}
H.~Li and C.~E. Saavedra, ``Linearization of active downconversion mixers at the if using feedforward cancellation,'' \emph{IEEE Transactions on Circuits and Systems I: Regular Papers}, vol.~66, no.~4, pp. 1620 -- 1631, 2019.

\bibitem{Garcıa-Naya2011}
J.~A. Garcıa-Naya, O.~Fresnedo, F.~J. Vazquez-Araujo, L.~Gonzalez-Lopez, M~andCastedo, and J.~Garcia-Frias, ``{Experimental Evaluation of Analog Joint Source-Channel Coding in Indoor Environments},'' \emph{in Proc. IEEE International Conference on Communications (ICC)}, 2011.

\bibitem{Zhang2014}
J.~Zhang, Y.~Wang, L.~Ding, and N.~Zhang, ``{Bit Error Probability of Spatial Modulation over Measured Indoor Channels},'' \emph{IEEE Transaction on Wireless Communication}, vol.~13, no.~3, pp. 1380--1387, 2014.

\bibitem{restuccia2020deepwierl}
F.~Restuccia and T.~Melodia, ``Deepwierl: Bringing deep reinforcement learning to the internet of self-adaptive things,'' in \emph{IEEE Conference on Computer Communications (INFOCOM)}, 2020, pp. 844--853.

\bibitem{XuCloudEEG}
X.~Wang, Y.~Zhu, Y.~Ha, M.~Qiu, and T.~Huan, ``An fpga-based cloud system for massive ecg data analysis,'' \emph{IEEE Transactions on Circuits and Systems II: Express Briefs}, vol.~64, no.~3, pp. 309 -- 313, 2017.

\end{thebibliography}

\vspace{-1.2cm}

\vspace{-1.3cm}

\vfill

\end{document}